\documentstyle[natbib,epsfig,longtable]{aipproc}

\newcommand{\ee}{{\rm e^+e^-}}
\newcommand{\mm}{ \mu^+\mu^-}
\newcommand{\lum}{{\rm \cal{L}}}  

\newcommand{\Lunits}{\,{\rm cm^{-2}.s^{-1}}}
\newcommand{\ECoM}{{\rm E_{CoM}}}

\begin{document}

\setcounter{secnumdepth}{2}    

\pagestyle{plain}
\footskip 1.5 cm

\title{Prospects for Colliders and Collider Physics to the 1 PeV
Energy Scale
\thanks{
To appear in Proc. HEMC'99 Workshop -- Studies on Colliders and
Collider Physics at the Highest Energies: Muon Colliders at 10 TeV to
100 TeV; Montauk, NY, September 27-October 1, 1999,
web page http://pubweb.bnl.gov/people/bking/heshop.
This work was performed under the auspices of
the U.S. Department of Energy under contract no. DE-AC02-98CH10886.}
}

\author{Bruce J. King}
\address{Brookhaven National Laboratory\\
email: bking@bnl.gov\\
web page: http://pubweb.bnl.gov/people/bking}
\maketitle

\begin{abstract}

  A review is given of the prospects for future colliders and collider
physics at the energy frontier. A proof-of-plausibility scenario is
presented for maximizing our progress in elementary particle physics by
extending the energy reach of hadron and lepton colliders as quickly and
economically as might be technically and financially feasible. The scenario
comprises 5 colliders beyond the LHC -- one each of $\ee$ and hadron
colliders and three $\mm$ colliders -- and  is able to hold to the
historical rate of progress in the log-energy reach of hadron and lepton
colliders, reaching the 1 PeV constituent mass scale by the early 2040's.
The technical and fiscal requirements for the feasibility of the scenario
are assessed and relevant long-term R\&D projects are identified.
Considerations of both cost and logistics seem to strongly favor housing
most or all of the colliders in the scenario in a new world
high energy physics laboratory.
\end{abstract}

\section{Introduction}
\label{sec:intro}


  No clear-cut consensus currently exists on the best long-term strategy
for experimentation in high energy physics (HEP) over the next 50 years.
This paper puts the case for continuing to aggressively raise the frontier
energy reach of both hadron and lepton colliders. It is argued that a
continuation at the historical rate of progress in the log-energy reach
of colliders is plausible and would provide us with outstanding prospects
for deepening our understanding of the elementary entities and organizing
principles of our physical Universe.

  In order to demonstrate the possible feasibility of such a push to higher
collider energies, a proof-of-plausibility scenario is presented for future
colliders that would continue at the historical pace in log-energy reach
and would, by about the year 2040, attain a constituent energy reach of 1 PeV
(i.e. 1000 TeV).

  The proof-of-plausibility scenario is only one choice from a parameter
space of plausible scenarios that might advance the energy reach of colliders
at the historical pace and, even if viable, no claim
is made that it is in any sense optimal.
Instead, it is intended as a spur to constructive criticism and future
research that will lead
to its refinement and to alternative scenarios.
Any such discussions will help us to assess the future prospects of HEP
and to identify long-term R\&D needs. In turn, this will enable the field
to make more informed planning decisions towards our long-term future.

  The paper begins with motivational background on the essential role of
past and future colliders for our understanding of elementary particle
physics. It then reviews the technical challenges of energy frontier
colliders and presents, and then evaluates, the aforementioned
proof-of-plausibility scenario.

  In more detail, the paper is organized as follows.
     Section~\ref{sec:motive} provides
a brief historical review of the impressive historical gains in physical
understanding from past accelerator experiments and then turns to an outline
of the heady physics goals for future colliding accelerators.
     Section~\ref{sec:wishlist} gives a wish-list, motivated by physics
considerations, for the technical specifications and capabilities of
future colliders. These include increased energy (mainly), specification
of the physics requirements for luminosity, and the physics advantages in
being able to study more than one type of projectile collisions.
     Section~\ref{sec:Livingston} reviews the rate of historical progress
in the energy reach of colliders as characterized by the famous
Livingston plot. It then introduces the proof-of-plausibility scenario
as an extension to the Livingston plot.
     The technical challenges and potential energy reaches
future $\ee$, hadron and $\mm$ colliders are briefly assessed in
sections~\ref{sec:electron} through~\ref{sec:muon}, respectively, and
justifications for the specific collider parameter choices of the
proof-of-plausibility scenario are embedded in the more general
discussions of these sections.
The scenario as a whole is then assessed in section~\ref{sec:scenario},
before concluding with a summary of the issues highlighted by the paper.


\section{Colliders at the Energy Frontier are Indispensable}
\label{sec:motive}

 The continuing motivation for colliding accelerators (colliders) is to
explore the most fundamental mysteries of the natural Universe: what are
its fundamental entities and organizing principles ? What is the nature
of space and time ? How did the Universe originate and evolve ?

 Accelerators provide us with experimental insights on our Universe that
could not plausibly be obtained in other ways and, to quote Harvard theorist
Sidney R. Coleman~\cite{Sci Am}, ``Experiment is the source of imagination.
All the philosophers in the world thinking for thousands of years couldn't
come up with quantum mechanics''. The properties and spectrum of elementary
particles have been no less hidden from theoretical understanding than was
quantum mechanics.

 This section reviews the central historical importance of accelerators
in uncovering what is known today as the Standard Model of elementary
particles. It then turns to our future aspirations for a yet deeper
understanding of the elemental entities and physical principles of our
Universe. In our most optimistic hopes, this might ultimately be described
by a complete and logically self-consistent ``theory of everything''.

\subsection{The Historical Importance of Accelerators}
\label{subsec:motive_past}

  The past fifty years of experiments at accelerators have lead to
remarkable progress in our understanding of the elementary processes
and building blocks of our physical Universe. We discuss, in turn, the
new insights gained on photons and on the building blocks of everyday
matter, and then briefly summarize our current state of knowledge as
encapsulated in the Standard Model of elementary particles.

\subsubsection{A Context for Understanding Photons}

  Surprisingly, accelerators have greatly expanded our understanding of
the multiple roles that photons play in the make-up and runnings of
our Universe. It is manifestly obvious that these important insights could
not have been attained without accelerator experiments -- and this despite
the fact that the
photons themselves are all around us and can be studied in many ways.

  In a famous quote~\cite{Einstein}, Albert Einstein once
confessed that ``All the fifty
years of conscious brooding have brought me no closer to the answer to
the question, ``what are light quanta?''''.
While we certainly still can't claim full understanding,
discoveries at accelerators have at least moved us towards a context and
framework for understanding the photon that even Einstein could not have
suspected. Far from being an isolated entity, the photon has massive
siblings -- known as the $Z^0$ and $W^\pm$ -- that have been produced
and studied at colliders, with observed masses of ${\rm M_Z=91.2\:GeV/c^2}$
and ${\rm M_W=80.4\:GeV/c^2}$, respectively. Their close relationship to
the photon has been well established from their observed properties, and
the interactions and relative couplings of
the photon, Z and W to other elementary particles
are now precisely specified by a theory, known as the electroweak theory,
that unites the electromagnetic and weak interactions.
(Experiments at upcoming colliders may further expand the scope of this
theoretical framework to include the mechanism for generating all
mass, as will be presented in the following
subsection~\ref{subsec:motive_future}.)

  In stark contrast to the photon itself, the large masses and immediate
decays of its sibling Z and W bosons clearly preclude their direct
study outside collider experiments. However, their effects are still seen
in our everyday world since it is their interactions that turn out to be
ultimately responsible for many radioactive decays -- an everyday phenomenon
that has been made a little less mysterious through the understanding
provided by the electroweak theory.

\subsubsection{Understanding Matter}

  Besides photons, our understanding of the building blocks of everyday
matter has been revolutionized by accelerators. Electrons have passed
progressively more stringent experimental tests of whether they are indeed
point particles. On the other hand, protons and neutrons have been
exposed as being composite entities rather than point particles.
Experiments at accelerators  have found them to be composed of hitherto
unknown elementary particles: up-type, down-type and a few strange-type
quarks bound together by gluons.

\subsubsection{A Periodic Table of Elementary Particles}

  Accelerator experiments and, to a lesser extent, cosmic ray experiments
have also shown us convincingly that the limitation of everyday matter to
electrons, protons and neutrons is merely an ``accident'' of our low energy
environment. Heavier forms of matter exist but cosmic ray interactions are
the only naturally occurring process on Earth to supply the energy densities
required to produce them. The few such particles that are produced
sporadically by cosmic rays then decay almost instantaneously (sometimes
in cascades of several stages) down to the familiar everyday particles --
photons, electrons, protons and neutrons -- plus the ghost-like neutrinos
that are also all around us but are almost undetectable. Accelerator
experiments are the only place where these particles can be systematically
studied.

  The list of additions to our everyday matter is impressive.
Besides the aforementioned W and Z bosons, the electron has been found
to have heavier siblings -- the muon and the tau particle, and each of these
has a neutral counterpart known as a neutrino that interacts so feebly that
it is difficult to detect. Hundreds of quarks-plus-gluons bound states besides
the proton and neutron have also been discovered, including bound states
containing other, heavier quarks than the up and down quark -- the so-called
second generation strange and charm quarks and the third generation bottom
(or beauty) and top quarks.

\begin{figure}[t!] %
\centering
\includegraphics[height=2.4in,width=2.8in]{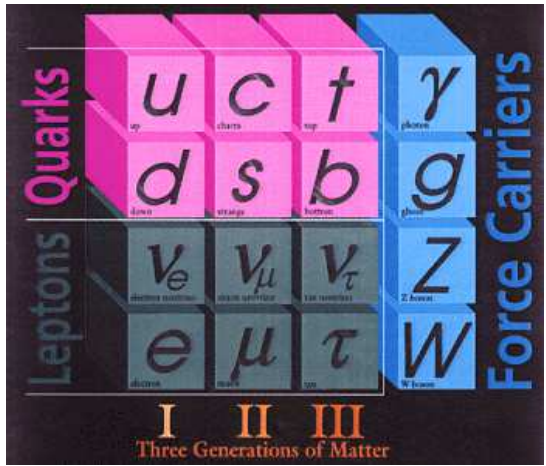}
\caption{
The known elementary particles. Illustration reproduced from
reference~\cite{FNALweb}.
}
\label{particlechart}
\end{figure}

  The new elementary particles fill out a veritable periodic table, which
is shown in figure~\ref{particlechart}. Although smaller than the more
familiar periodic table of the elements, the structural patterns are more
complicated. The grouping of the particles in figure~\ref{particlechart}
reflects the particles' properties and the interactions they participate in.
These properties and interactions are all well described by the so-called
Standard Model of elementary particles~\cite{SM} , which will be
discussed further in the following subsection.

  To summarize our past progress, accelerator experiments have already
revolutionized our understanding of the elementary building blocks and
interactions in the Universe around us. They have led to the standard
model of elementary particles and have exposed the intrinsic naivete
of any pre-accelerator picture of the Universe that had the proton,
neutron, electron and photons as the sole elementary particles. However, they
have also highlighted our continuing naivete. Further collider experiments
will be essential to achieve a more satisfying level of understanding, as
will now be explained.

\subsection{Heady Physics Goals for Future Colliders}
\label{subsec:motive_future}

\subsubsection{The Standard Model is only a Stop-Gap Theory}

  Even though the crowning achievement of the past 50 years of HEP has
been the construction of the Standard Model, our intellectual goals for its future
are very much more wide-reaching than simply filling out the ``periodic
table'' of figure~\ref{particlechart} and further detailing its properties.
For, despite its predictive power for existing HEP experiments, we know
the Standard Model to be no more than a stop-gap theory with a limited domain of
applicability. It is phenomenological
rather than fully predictive, incorporating 19 free parameters that need
to be determined by experiment. Even more damning, it becomes logically
inconsistent when we try to extrapolate it to experimentally inaccessible
energy scales.

  Instead, the quest is for a deeper knowledge of where the Standard Model comes from
and also for understanding its connection to the existence and bulk properties
of the Universe such as its preponderance of matter over antimatter and its
gravitationally-curved 3+1 dimensional space-time structure.

\subsubsection{Strategies for Advancing beyond the Standard Model}

  This paper deals largely with the paramount importance of energy
frontier colliders as discovery machines to further uncover the secrets
of elementary particles and, hence, to learn more about the physical
foundations of the Universe. This emphasis on energy reach
is further justified in section~\ref{subsec:wishlist_E}.

  Apart from the approach emphasized here, it is worthwhile to briefly
mention some other of the well-developed alternative or complementary
strategies for examining these questions at the current and next
generations of colliders. More detailed expositions can be found
elsewhere in these proceedings~\cite{Dawson, Willis, Parsa}.

  Very briefly, the origin of all mass in the Universe is hypothesized,
in the Standard Model, to be intimately tied in with an as yet undiscovered particle
known as the Higgs boson. This is a hypothesized extension to the
electroweak theory that was discussed previously -- that is to say,
our context for
understanding photons could further lead to a context for understanding
the generation of all mass! As such, it provides a beautiful example
of how experimental advances in elementary particle physics might
build up a level of understanding of our Universe that would have be
otherwise unattainable. The LHC has been optimized to search for this
Higgs particle and proposed TeV-scale linear colliders have been
optimized for follow-up precision studies of such a particle, if it
exists. (Note, however, that both the LHC and $\ee$ colliders will
anyway be at the energy frontier, so this is merely a shading of the
approach emphasized in this paper rather than a true alternative
strategy.)

  As another thread~\cite{Willis}, colliders have the potential to reproduce
and study the extreme energy densities that existed in the first moments after
the Universe formed in the hypothesized ``Big Bang''. Collisions of heavy
ions -- at the RHIC collider, for example -- will provide the largest
volumes under such conditions, even though these volumes are admittedly
still miniscule.

  Related to this, it has been hypothesized that the observed dominance of
matter over antimatter in the observable Universe
might have originated from the properties of exotic heavy particles
and their anti-particles that could have been routinely produced in the
earliest moments of the Universe and whose interactions could have involved
large matter-antimatter asymmetries -- this is known as CP violation in the
obscuring lingo of the field. We may get an experimental handle on such
possibilities from experiments at B factory colliders and elsewhere that
study the effects of CP violation.

\subsubsection{The ``Theory of Everything''}

 Ultimately, the ``holy grail'' of high energy physics is to advance
from the Standard Model to the hypothesized ``Theory of Everything'' that describes
and explains the elementary entities, structure and organizing principles
of our Universe and is predictive -- at least in principle, even if not
calculationally -- for any experiment we could conceive of that
involves elementary particles.

 The nature of the Theory of Everything
and any possible intermediate levels of
understanding towards the Theory of Everything are the
subjects for current speculation
and future discoveries at accelerators, coupled to theoretical breakthroughs
in interpreting these discoveries. Speculation on the elementary particle
physics phenomena we might find at future accelerators is helpful in
stimulating theoretical progress towards
the Theory of Everything and also for the design
of the future colliders and their experiments. Several example scenarios
for what we could find at many-TeV muon colliders have been presented in
these proceedings~\cite{Berger,Lane,Rizzo}, along with a very helpful
classification scheme for the possibilities~\cite{Lykken}.

 The Cambridge theorist and physics popularizer Steven Hawking gives the
field a 50\% chance~\cite{Hawking} of attaining the Theory of Everything within the next
20 years. This is the most optimistic assessment I am aware of and had
many of the theorists at this workshop shaking their heads. At the opposite
extreme, it is even logically possible that a unified understanding of
the Universe is simply beyond human intellect or, as Steven Weinberg put
it, like trying to explain Newtonian mechanics to your dog!

 In any case, we know it will not be {\em easy} to reach a complete
understanding of the physical foundations for our Universe, and collider
experiments at the energy frontier will presumably be our main experimental
tool in this heady quest.

\section{A Wish-list for the Parameters of Future Colliders}
\label{sec:wishlist}

  This section begins by stressing the paramount importance of energy
reach in determining the physics potential of future colliders. (It
should be acknowledged that the viewpoint expressed here has certainly
been influenced by other presentations at the HEMC'99 workshop and
contributions to these proceedings, particularly that of
Samios~\cite{Samios} and the summary presentation by
Willis~\cite{Willis}.)

 The auxiliary requirements for, and benefits of, adequate luminosity and
utilizing a variety of colliding projectiles are then discussed
in subsections~\ref{subsec:wishlist_lum}
and~\ref{subsec:wishlist_proj}. Beam polarization
and clean event reconstruction are other relevant experimental
capabilities whose discussion is left to elsewhere in these proceedings,
in references~\cite{Heusch} and~\cite{Kahn,Rehak}, respectively.

\subsection{The Paramount Importance of Energy Reach}
\label{subsec:wishlist_E}

\subsubsection{Energy Reach versus Less Direct Experimental Strategies}

  Some specific experimental strategies for extending our knowledge
beyond the Standard Model were discussed in the preceding section. However, the
only way we can {\em directly} examine an energy scale is by cranking up
the energy of our colliders to reach that energy scale. This will
then allow us to observe any exotic particles or even more complicated
entities~\cite{Lykken} that might exist at that energy scale, whether
or not they had been previously forecast on theoretical grounds. Hence,
a direct frontal assault on the collider energy frontier is intrinsically
more powerful and more likely to result in major break-through discoveries
than are alternative, more indirect experimental approaches.

\subsubsection{Energy Frontier Colliders Can Also Do Lower Energy Physics}
\label{sss:wishlist_energy_lowerE}

  In weighing the balance for future frontier machines versus lower
energy colliders it should be borne in mind that, besides their primary
mission of discovery, frontier machines can also do well at studying
lower energy processes -- often even better than at dedicated lower-energy
facilities.

  As an example, the LHC, with $E_{CoM}=14$ TeV, will be one of the
best places to do studies with B ($M \simeq 5\: GeV/c^2$) and charm
($M \simeq 1.7\: {\rm GeV/c^2}$) particles. Even lepton colliders have
the general property that lower mass particles are produced in higher
order processes and in the decays of heavier particles.

 A currently relevant example with lepton colliders is given by
collider parameter sets for the 10 TeV and 100 TeV muon colliders
that were studied at this workshop~\cite{hemc99specs}. The specified
luminosities would correspond to the production of more than $10^7$
Standard Model Higgs particles if these existed at the 100 GeV mass scale.
This would be orders of magnitude more events than at any of the lower
energy electron or muon colliders that have been proposed with the
principal goal of studying such a Higgs. (Admittedly, less precise
event reconstruction may somewhat dilute the statistical advantage.)
Therefore, at least some aspects of any Higgs particle, such as rare
decay modes, might be better addressed at frontier colliders than at
dedicated Higgs machines operating at the few hundred GeV energy scale.

 Besides examples using future colliders,
the case can also be made in a historical
context, as now follows.

\subsubsection{An Alternative History: The Standard Model could have
  been Reconstructed from Today's Energy Frontier Experiments Alone}

 We now consider a historical ``what if'' question that highlights both
the paramount importance of energy reach and the ability of energy frontier
colliders to perform analyses concerning lower energy scales.

  Consider the state of elementary particle physics a half century ago,
in 1950. The positron
(1933), muon (1937) and pion (1947) had been discovered in cosmic ray
experiments, following up on the discovery of the neutron (1932) and the
inferred existence of the neutrino from beta decay spectra (1932-3).
The historical gedanken experiment is to imagine that, instead of the
newly commissioned 184-in synchrocyclotron at Berkeley, which could produce
pions, the HEP community of 1950 had been immediately gifted with today's
energy-frontier hadron and lepton colliders -- the 1.8 TeV Tevatron
proton-proton collider and the 90--200 GeV LEP electron-positron collider
-- along with the technology for their modern-day general purpose
collider experiments.

  We can then ask the following question: how much of today's current
understanding of elementary particles (i.e. the Standard Model) would have
been promptly reconstructed from the data and what, if anything, would have
been missed ?

  It can be argued that the basic structure of the Standard Model would have been
quickly recovered -- either in its entirety or nearly so -- since the
Tevatron and LEP see evidence for all of the particles in
table~\ref{particlechart} (redundantly, in most cases) and provide
measurements of their interactions and couplings.

 In more detail, the copious production of W's and Z's would quickly
arrive at the electroweak theory that was mentioned previously in
section~\ref{subsec:motive_past}. Knowledge of the strong interaction
and of the point-like quarks and gluons it acts on would also come easily,
from observations of the ``jettiness'' of hadronic events at both
colliders and from other evidence, and these event signatures would
also show immediately that the Tevatron's proton projectiles were
composed of these quarks and gluons.

  Probably the last piece of the Standard Model structure to be experimentally
established in this scenario would be the complex phase in the CKM
quark mixing matrix that accounts for CP violation.
%
%
The energy frontier collider experiments are poorly optimized for
observing the small effects of CP violation in kaon decays where
this phenomenon was experimentally discovered, and the alternative
of experimental
evidence for CP violation in B decays still has only marginal statistical
significance at both LEP~\cite{CPviolOPAL} and the Tevatron~\cite{CPviolCDF}.

  Even though CP violation is the part of the Standard Model least suited for study
at the Tevatron and LEP, their data would certainly still provide the CKM
matrix as a theoretical construct and would show it to be non-diagonal.
From there, theoretical conjecture on a possible complex phase would
be natural and this could well lead to a re-optimized detector (e.g.
similar to the B-TeV detector that has been proposed for the Tevatron)
that could follow up with more definitive measurements of CP violation
to complete the picture of the Standard Model.

  To summarize, the outcome of the above gedanken experiment reinforces
the previous conclusions of this section by demonstrating that today's
energy frontier colliders can quickly provide access to all of the
elementary particle physics structure that we are aware of from our 50
years of historical progress.

\subsection{Desirable Luminosities and their Scaling with Energy}
\label{subsec:wishlist_lum}

 The luminosity of future high energy colliders is the machine
parameter that is second in importance only to energy reach.

  A rule of thumb for hadron colliders that came into prominence
in the 1980's is that the physics gain from a factor of
10 in a hadron collider's luminosity corresponds roughly to factor
of 2 in energy reach for hadron colliders. The possibilities for
such a trade-off are presumably more limited for the point-like
projectiles of lepton colliders, where $\ECoM$ gives a more
precise measure of the discovery energy reach.

 Probably the best way to define the luminosity goals for energy
frontier colliders is that the luminosity should be sufficient
to gather good statistical samples for the study of any elementary
particles existing at the energy scale, $E \le \ECoM$.

  This definition raises a conundrum for discovery colliders at
the energy frontier: the number of events is given by the product
of the production cross section for the particle, $\sigma$, and
the time integral of the luminosity,
$\lum$:
\begin{equation}
 {\rm no.\: events} = \sigma \int \lum \, {\rm dt};
\end{equation}
yet how can we predict the cross sections for unknown particles ?

 Fortunately, it is common knowledge that very approximate upper limits
for production cross sections as a function of collider energy can be
guessed at just from general considerations of relativistic quantum
mechanics. We now give a version of the type of hand-waving argument
that makes this connection. This argument works for the point-like
projectiles of lepton colliders at any chosen $\ECoM$. The very approximate
luminosity specifications that result could arguably be extended to hadron
colliders by replacing $\ECoM$ with the equivalent energy reach,
$E_{reach}^{pp} \simeq \ECoM /6$, for the collisions of the quark
and gluon sub-components of protons.
(See section~\ref{sec:Livingston} for further discussion on the
parameter $E_{reach}^{pp}$.)

  The venerated Heisenberg uncertainty principle of non-relativistic
quantum mechanics,
\begin{equation}
\Delta p\, \Delta x \ge \frac{\hbar}{2},
    \label{nonrelHeisenberg}
\end{equation}
can, for elementary particles, be recast into the very approximate
relativistic form
\begin{equation}
\Delta E\, \Delta x \sim \hbar c,
    \label{relHeisenberg}
\end{equation}
where $\Delta p$ is the momentum spread of a particle's wave-function,
$\Delta x$ is the position spread, $\hbar$ is the reduced Plank's
constant, $\Delta E$ can be considered as the energy scale of an
interaction and $\Delta x$ gives the corresponding spatial extent,
and the conversion from equation~\ref{nonrelHeisenberg}
to equation~\ref{relHeisenberg} uses the approximate ultra-relativistic
relation $E \simeq pc$ that neglects the incoming particle masses.
The cross sectional area over which the interaction can occur
will be of order the square of the spatial extent of the interaction,
roughly $(\Delta x)^2$, so the maximum cross section for a given
center-of-mass energy will be roughly:
\begin{equation}
\sigma_{max} \sim (\Delta x)^2 \sim
     \left( \frac{\hbar c}{E_{CoM}} \right)^2,
  \label{QMxsecexpress}
\end{equation}
or, numerically,
\begin{equation}
\sigma_{max} [pbarn] \sim \frac{400}{(E_{CoM}[TeV])^2},
  \label{QMxsec}
\end{equation}
where units are given in square brackets in this equation and throughout
this paper, and 1 picobarn (pbarn) is $10^{-12}$ barn or
$10^{-36}\:{\rm cm^2}$.

 The crude estimate of equation~\ref{QMxsec} actually does surprisingly well
at predicting the largest cross section at today's 100-GeV-scale $\ee$
colliders: it predicts $\sigma_{max} \sim 50$ nbarn at the energy of the Z
pole, 91.18 GeV, which agrees well with the actual cross section of 38 nbarn.

  Apart from the Z resonance, however, most cross sections for point-like
interactions have been observed to fall several orders of magnitude below
the value of equation~\ref{QMxsec}. This can be explained away, in the
hand-waving spirit that the equation was derived, by saying that any coupling
suppressions arising from the detailed physics process will generally reduce
the probability of an interaction occurring for even the closest encounters.

  As an acknowledgment that large coupling suppressions are the norm
rather than the exception, the luminosities of lepton colliders are
commonly bench-marked to a process other than resonant Z production, namely,
to lepton-antilepton annihilations to fermions through photon exchange,
e.g.,
\begin{equation}
{\rm e^+e^-} \stackrel{\gamma}{\rightarrow} \mu^+\mu^-.
  \label{Rprocess}
\end{equation}
This has a cross section of:
\begin{equation}
\sigma_{R} [pbarn] = \frac{0.087}{(E_{CoM}[TeV])^2}.
  \label{Rxsec}
\end{equation}
(To be precise, this only gives accurate predictions for the cross
section at energies well below the Z resonance, at 91 GeV. At energies
above this, the cross section is substantially modified by interference
with the corresponding process involving Z exchange. Instead,
equation~\ref{Rxsec} is intended as the {\em definition} of a benchmark
cross section that can be used at all energies, as we now explain.)

 The inverse of the characteristic cross section of equation~\ref{Rxsec}
defines a unit of integrated luminosity known as a ``unit of R'' such
that a collider that collects one unit of R of integrated luminosity will
produce, on average, one event that has the cross section of
equation~\ref{Rxsec}.

  It was the guidance~\cite{Peskin} of SLAC theorist Michael Peskin that
the luminosity for this workshop's straw-man muon collider parameter sets
should, if possible, allow an accumulated inverse luminosity of 10 000
units of R. To convert this to an average luminosity it can be noted that
obtaining this integrated luminosity over $5 \times 10^7$ seconds of running
(five ``Snowmass accelerator years'') requires an average luminosity of:
\begin{equation}
\lum^{\rm desired} {\rm [cm^2.s^{-1}]} =
             {\rm  2.3 \times 10^{33} \times \left( E_{CoM}[TeV] \right)^2.}
      \label{Ldesired}
\end{equation}
This can be contrasted with the much more modest luminosity that would
be needed to acquire 10 000 events produced at the approximate maximum
cross section specified by equation~\ref{QMxsec}:
\begin{equation}
\lum^{\rm borderline} {\rm [cm^2.s^{-1}]} =
             {\rm  5 \times 10^{29} \times \left( E_{CoM}[TeV] \right)^2.}
    \label{Lborderline}
\end{equation}

 The straw-man parameters for both the 10 TeV and ``100 TeV ultra-cold beam''
examples met or even exceeded Peskin's request, each with 8700 units of
R per detector in a single year. On the other hand, the ``100 TeV evolutionary
extrapolation'' parameter set specified only 87 units of R per detector
per year. This reflects the escalating luminosity demands with $\ECoM$
due to the $1/\ECoM^2$ cross section scaling of
equations~\ref{QMxsec} and~\ref{Rxsec}.

\subsection{The Complementarity of Different Projectile Types}
\label{subsec:wishlist_proj}

  The different experimental
conditions and, particularly, the different interacting projectiles
of hadron and lepton colliders will generally lead to different
sensitivities for specific processes at the energy scale under
consideration, so the two types of colliders are also complementary
to a certain extent and there are advantages to operating both types
of machines. This complementarity also applies to the two types of lepton
colliders -- $\ee$ and $\mm$ colliders -- but to a lesser extent.

  There are also many other possibilities for the colliding projectiles
that will not be discussed further in this paper:
gamma-gamma collisions, heavy ion colliders, like-sign lepton
colliders~\cite{Heusch} (${\rm e^-e^-}$ and $\mu^-\mu^-$) and any one
of the several options that collide dissimilar projectiles. These options
all have some potential for complementary physics studies and should
be looked at further. However, it should be noted that several of them
are understood to be less suitable for exploring the energy frontier
for various reasons, a few of which are discussed elsewhere in these
proceedings~\cite{Telnovprojectiles}.

 In the past, energy frontier hadron colliders have been regarded more
as discovery machines while lepton colliders, following later but with
cleaner experimental conditions, have been considered mainly as follow-up
machines for precision studies. The following section reviews the history
of collider facilities that led to this assignment, as well as
introducing a speculated scenario for future colliders.

\section{The Livingston Plot for Progress in the Energy Reach of
Colliders --  Past, Present and Future}
\label{sec:Livingston}

\begin{figure}[t!] %
\centering
\includegraphics[height=3.0in,width=4.5in]{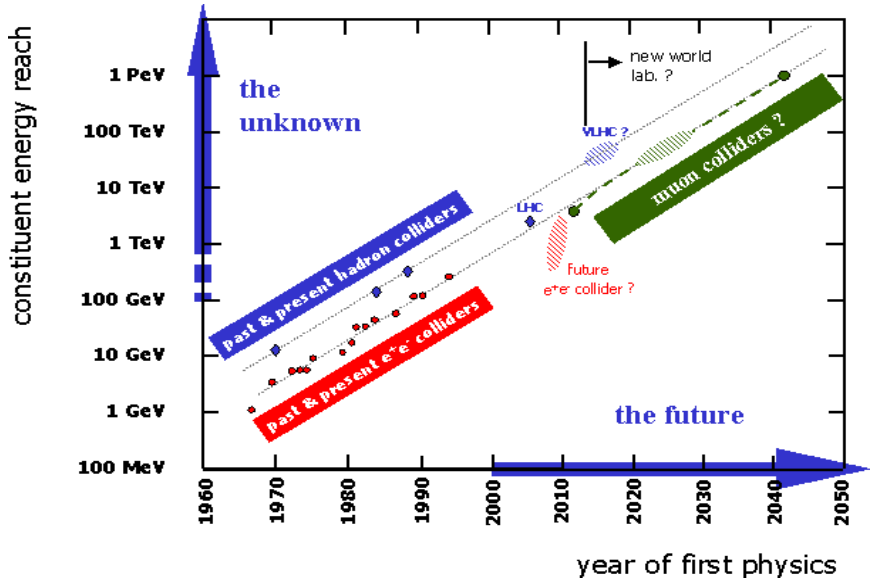}
\caption{The
Livingston plot showing the historical progress in the constituent
energy reach of lepton and hadron colliders.  Each point on the curve
represents a collider. A possible scenario for future colliders
to continue the exponential progress in hadron and lepton energy reach has
been added. The definitions and data points in the plot are discussed
further in the text, as is the scenario for future colliders.
}
\label{Livingston}
\end{figure}

\subsection{Presentation and Interpretation of the Plot}
\label{subsec:Livingston_plot}

  Figure~\ref{Livingston} is the famous Livingston plot showing the
historical exponential growth with time in the energy reach of both
lepton and hadron colliders. The data for past and present lepton and
hadron colliders has been taken from reference~\cite{NLC} and is discussed
and parameterized in the following subsection.

  The logarithmic energy scale in figure~\ref{Livingston} is physically
appropriate under the reasonable assumption that the underlying physical
importance of the mass spectrum will lie in the {\em ratios} of particle
masses as opposed to mass {\em differences}. As some confirmation of this
assumption, the masses of the known elementary particles do indeed fall
relatively evenly along a log-energy scale rather than being bunched at
the low energy end. To rephrase this in a way that might sound depressing
to accelerator builders, the past exponential progress in the
energy reach of colliders can be considered to have corresponded
to merely a steady (linear) rate of advance in their physics capabilities
since the logarithm of the energy rather
than the energy itself is the appropriate metric for assessing
the discovery reach of colliders.

 Some speculated future colliders beyond the LHC have been added in
to figure~\ref{Livingston}. In sum, they are intended to comprise
a straw-man proof-of-plausibility scenario to show
that a sufficiently motivated and
adequately funded HEP community may be able to continue constructing
accelerators that lie on or near the lepton and hadron Livingston curves
and that extend up to the PeV constituent energy scale
(where 1 PeV = 1000 TeV).
Discussion on this scenario occupies the final subsection in this section,
subsection~\ref{subsec:Livingston_PoP}, as well as
much of the remainder of this paper.

\subsection{Parameterizations for the Historical Progress in the Energy
      Reach of Hadron and Lepton Colliders}
\label{subsec:Livingston_para}

  The constituent energy reach for lepton colliders in figure~\ref{Livingston}
has been defined simply to be their center-of-mass energy,
\begin{equation}
E_{reach}^{lepton} \equiv E_{CoM}.
   \label{Elepton}
\end{equation}
The fact that protons are composite particles rather than point-like
elementary entities dilutes the constituent energy reach of hadron colliders
relative to lepton colliders by a factor that depends on both the
physics process and the collider luminosity.
The choice of dilution factor for past hadron machines was copied from
reference~\cite{NLC} and we have chosen the similar dilution factor
of 6 for future hadron colliders:
\begin{equation}
E_{reach}^{pp} \equiv E_{CoM}/6.
  \label{Epp} 
\end{equation}
For any given hadron collider, other estimates for the dilution factor may
differ by a factor of two or more from this choice, in either direction.
There is also a slight arbitrariness in some of the other data choices,
so the reader is warned that the details in Livingston plots may
vary from publication to publication.


 The two dashed lines drawn through the data points give parameterizations
for the constituent energy reach, $E_{reach}$, versus year of first
physics, Y, for lepton and hadron colliders. They have equations:
\begin{equation}
\log_{10}( E_{reach}^{lepton}[TeV] ) = (Y-2002) / 13
    \label{Ylepton}
\end{equation}
and
\begin{equation}
\log_{10}( E_{reach}^{pp}[TeV] ) = (Y-1994) / 13,
    \label{Yhadron}
\end{equation}
respectively and substituting
the energy dilution factor from equation~\ref{Epp} into
equation~\ref{Yhadron} gives the required CoM energy reach for future
hadron colliders:
\begin{equation}
\log_{10}( E_{CoM}^{pp}[TeV] ) = (Y-1984) / 13.
    \label{CoMhadron}
\end{equation}

  Equation~\ref{Ylepton} informs us that a decade of energy increase
in lepton colliders has historically occurred every 13 years. Proton
colliders have advanced at the same rate of progress as lepton colliders --
an energy decade every 13 years -- but have been about 8 years ahead of
lepton colliders in attaining a given constituent energy reach.

  As well as past and present colliders, figure~\ref{Livingston}
plots the planned 2005 completion date for the Large Hadron Collider
(the LHC, currently under construction) and a region representing
roughly the range of predicted or proposed turn-on dates and energies
for contemplated electron-positron colliders. Beyond these, later and
more speculative collider points at higher
energies have also been added to figure~\ref{Livingston}.
A straw-man ``target'' region for a Very Large Hadron Collider
(VLHC) is shown with an energy range of 200--400 TeV corresponding
to first physics dates of
2014-8. Three muon collider points/regions are also shown,
extending up to the lepton Livingston curve and with CoM energies up to 1 PeV.

 It can be noted that the cancelation of the SSC collider has hampered
progress in the energy reach of hadron colliders, as is also stressed
elsewhere in these proceedings~\cite{Samios}. The SSC would have been above
the Livingston curve for hadrons if it had already been built at its
design energy of $E_{CoM}=40$ TeV. Instead, it can be seen from
figure~\ref{Livingston} that the LHC has already fallen below
the Livingston curve; to have stayed on the curve,
equations~\ref{Epp} and~\ref{Yhadron} specify that it would have needed to
have either turned on last year (1999) or, for its planned 2005 turn-on,
had an $E_{CoM}$ of 42 TeV rather than 14 TeV.


\subsection{Introducing the Proof-of-Plausibility Scenario for Future
                                                              Colliders}
\label{subsec:Livingston_PoP}

\begin{table}[htb!]
\caption{
A time-line for energy frontier muon colliders that would
hold to the historical rate of progress in the highest
collision energies attained with lepton colliders.
}
\begin{center}
\begin{tabular}{ccl}
year & $E_{CoM}$ \&  type &  description \\
\hline
2009-11 & 0.4-3 TeV $\ee$ & EW-scale/energy frontier linear collider    \\
2012    & 4 TeV $\mm$     & lab. site-filler with low beam current      \\
2014-18 & 200-400 TeV pp  & 400 km circum. is boundary for new lab.     \\
2021-8  & 30-100 TeV $\mm$ & high luminosity collider in new world lab.  \\
2041    & 1 PeV $\mm$     & linear collider with electron drive beams   \\
\end{tabular}
\label{timeline}
\end{center}
\end{table}


  Table~\ref{timeline} summarizes the collider type, energy reach
and year of first operation for the colliders in the proof-of-plausibility
scenario to extend the Livingston plot beyond the LHC.

  The scenario is economical in requiring only 5 colliders to reach all
the way up to the 1 PeV constituent energy scale -- one each of $\ee$ and
hadron colliders and three $\mm$ colliders. The correspondingly large
leaps in energy continue the necessary trend that was first set by the SSC,
with $\ECoM=40$ TeV, which improves
by more than a factor of 20 from the existing
energy frontier at $\ECoM=1.8$ TeV. (The SSC ended up being too expensive,
but this should not be interpreted as a fundamental flaw in the concept.)
The increasingly large energy jumps are dictated by the rising cost of
each successive machine. They are also desirable to give each new machine
a big enough window for a good chance to discover new physical phenomena
or, at minimum,
to rule out a big enough energy range for this to have theoretical
significance.

  In order to limit the number of colliders in table~\ref{timeline} to five,
the energies of the colliders have all been pushed up as far as was
considered practical towards, in each case, a natural and rather sharply
defined technological bound. 
For the first collider in table~\ref{timeline} -- a TeV-scale $\ee$
collider -- the fundamental energy bound comes from the sharp rise with
energy in the interactions between TeV-scale electrons and electromagnetic
fields, as will be discussed in section~\ref{sec:electron}.

  Only one extra hadron collider was included in the scenario
since the energy scale of the LHC,
with $\ECoM=14$ TeV, will already be enough energy to probably
warrant jumping in a single step
up to the few hundred TeV energy bound that results
from synchrotron radiation.
The limit and the 200-400 TeV hadron collider will be discussed
in section~\ref{sec:hadron}.

  The 3 muon colliders all butt up against
different technological energy limits, as will be further addressed in
subsection~\ref{subsec:muon_PoP}. The 4 TeV muon collider is
about at the size limit for fitting on a laboratory site and, more
importantly, is fighting against the energy-cubed rise in the off-site
neutrino radiation hazard for populated locations. The middle muon
collider pushes the synchrotron radiation energy limit for circular
muon colliders, which is probably in the range of 30-100 TeV. Finally,
1 PeV is a limiting energy scale for any linear muon colliders that
can't call on exotic acceleration schemes to obtain accelerating gradients
that greatly exceed today's technological bounds.

  While the scenario looks plausible at first sight, it is important
to stress that this is only a first presentation, and the scenario's
feasibility or otherwise should be better established with further studies.
Also, no claim is made that the scenario is in any way optimal and it
is hoped that follow-up studies and technical progress will lead to
refinements and also to alternative scenarios. This would give us more options
and possibilities to progress with foresight
towards a productive future for HEP.

%
%
%

\section{Future Electron-Positron Colliders}
\label{sec:electron}

\subsection{Energy Limitations on $\ee$ Colliders}
\label{subsec:electron_limits}

The fundamental energy limitation for electron-positron colliders is
the projectile itself. At only 1/1837 of the proton mass, electrons are
relativistic enough at TeV energy scales and beyond to be overly sensitive
to electromagnetic fields. They become difficult to bend, focus or collide
because of their excessive tendency to throw off photons due to synchrotron
radiation in the
magnetic fields of bending and focusing magnets and then to beamstrahlung
in the electromagnetic fields of the oncoming bunch at collision.

  Even to advance to the TeV scale, these difficulties have already required
the major technology shift from circular colliders to single pass linear
colliders. Synchrotron radiation scales as the fourth power of the beam
energy (or as the third power, for some parameters),
so further technological innovations would clearly be needed to
advance to still higher energies. To solve each of the three classes of
problems, we would need all three of:
\begin{enumerate}
   \item  acceleration that doesn't have a prohibitive linear cost
coefficient with energy
   \item  focusing to collision that doesn't induce excessive synchrotron
radiation
   \item  some way of damping out the beamstrahlung at collision.
\end{enumerate}

 Research is underway to attempt solutions to each of these three problems,
which have arguably been presented in order of increasing difficulty.
Potential solutions include electron drive beams for acceleration, plasma
focusing or focusing induced by auxiliary beams, and 4-beam schemes that
also use additional beams to partially cancel out the electromagnetic
fields at collision.

%
%

\subsection{This Scenario: A Single TeV-Scale Electron-Positron Collider}
\label{subsec:electron_scenario}

  Planning for proposed TeV-scale $\ee$ colliders is far enough along
that the shaded region for $\ee$ in figure~\ref{Livingston} simply
represents the very approximate range of energies and turn-on dates
for proposed machines.

  It can be seen that the shaded region extends well below the Livingston
curve. This reflects the technological problems associated with energy
that were discussed in the preceding subsection and,
somewhat related to this, the
widespread support for the alternative motive of studying identified
physics processes -- such as production of top quarks or Higgs particles
in the 100-200 GeV mass range, if they exist -- rather than a dedicated
thrust towards the energy frontier.

  The scenario assumes that exactly one TeV-scale $\ee$ collider will
be constructed before the mantle of frontier energy lepton collider 
is passed on to muon colliders. Besides its own physics value, this
$\ee$ collider would continue to establish the technology needed for
the later muon colliders and, in particular, for the assumed 1 PeV
muon collider that completes the scenario. The technological overlap
between the two types of lepton colliders is discussed further in
section~\ref{subsec:scenario_tech}.


\section{Future Hadron Colliders for the Energy Frontier}
\label{sec:hadron}

  This section discusses current research towards future
hadron colliders and assesses the constraints on
their ultimate potential energy
reach. As a particular constraint on their energy reach, it is
explained why linear hadron colliders will probably never be viable.
Finally, the 200-400 TeV pp collider in the 
proof-of-plausibility scenario is discussed.

\subsection{Current Research Towards Future Hadron Colliders}
\label{subsec:hadron_res}

  Beyond the LHC, research is underway for a follow-up hadron collider,
usually referred to as the ``Very Large Hadron Collider'' (VLHC),
whose goal would be to reach substantially higher energies than the LHC
without being much more expensive. The U.S. research efforts towards a
VLHC are conveniently summarized in the annual report of the  U.S. Steering
Committee for a VLHC~\cite{VLHC}.

  Much of the VLHC research involves magnet design because the bending
magnets for the collider storage ring are assumed to be a large, if not
dominant, component of the cost for a VLHC.
(For comparison, dipole magnet
costs for the collider ring of the SSC were budgeted to comprise
roughly 25\% of the total cost.) The current
emphasis on magnet R\&D is largely divided
between the design of low field (2 Tesla)
superferric magnets and very high field (greater than 9 Tesla) magnets.

 The low field superferric magnet designs might be very cheap
to construct. As one of the challenges for this
option, there is concern that the attainable
luminosities for the low field superferric option
may be limited due to beam instabilities caused by the combined
effects of the small magnet
apertures, the large collider circumference and the lack of synchrotron
radiation damping. The potential lack of stability and tune-ability for
such simple magnets is also a concern.

 A major motivation given for very high field magnets,
as opposed to intermediate field magnets, is their potential to cause
a beneficial level of synchrotron radiation damping of the beam at an
assumed beam energy of 50 TeV. The radiation damping is higher for high
field magnets because the fractional energy loss per
turn, ${\rm \frac{\Delta E}{\ECoM} }$ due to
synchrotron radiation scales with $\ECoM$ and with the average bending
magnetic field, B, according to:
\begin{equation}
{\rm \frac{\Delta E}{\ECoM} } \propto {\rm \ECoM^2 \times B }.
   \label{synchscaling}
\end{equation}
Very high field magnets cannot use the niobium-titanium superconductor
that has been used in all collider magnets to date since this conductor has
an impractically low critical current at magnetic fields above 9 Tesla.
Other superconductor materials
must be used, such as niobium-tin or the new high-$T_C$ superconductors,
and these are -- at least at present -- considerably more expensive
than niobium-titanium and have inferior mechanical properties.

  Regardless of progress in superconducting materials, the mechanical stresses
in magnets scale as the square of the magnetic field and will always conspire
to raise the cost per unit length of high field magnets relative to those
at lower fields. On the other hand, a more relevant yardstick than cost
per unit length is the cost per Tesla-meter since a collider ring at a
given energy will need a fixed number of Tesla-meters to bend the beams
in a circle. This favors higher field magnets so long as the cost per meter
increases less than proportionally to the field strength. Also, higher
fields should reap further cost savings from the consequent reduction in
tunnel length. Because of these trade-offs, the field strength for
superconducting magnets that would give the optimal cost for a future
hadron collider is not at all well established. (However, see the
discussion in subsection~\ref{subsec:hadron_PoP} regarding a study
by Willen~\cite{Willen} that uses {\em today's} superconducting magnet
technology.)

  More generally, it appears that the cost and technology optimizations
used for the VLHC could usefully be extended to include varying the energy
of future hadron colliders away from the VLHC's assumed $\ECoM=100$ TeV,
as will be addressed in subsection~\ref{subsec:hadron_PoP}.

\subsection{The Ultimate Energy Reach for Hadron Colliders}
\label{subsec:hadron_limits}


%

\subsubsection{Limits from Synchrotron Radiation}

  For the VLHC studies at $\ECoM = 100$ TeV, the beneficial damping
effects of synchrotron radiation must already be balanced against the
problems it causes. Given the strong power-law rise in synchrotron radiation
with energy, it can be surmised that synchrotron radiation should lead to
a fairly sharp technical cut-off in the viability of hadron
colliders by the few hundred TeV range.
More quantitatively, the power radiated due to synchrotron radiation,
${\rm P_{synch}}$, is given approximately by:
\begin{equation}
{\rm P_{synch}[MW]} \simeq 0.6 \times {\rm I[A] \times B[T] \times
                             \left( \ECoM [100\:TeV] \right) ^3 },
        \label{ppsynch}
\end{equation}
where I is the average current in each proton beam and B is the bending
magnetic field which, for this equation,
is simplistically assumed to be constant around a circular collider ring.

\subsubsection{Constraints from the Experimental Environment}

  Probably the biggest technical challenges for the SSC and LHC came, not
from the colliders themselves, but from the extreme operating conditions
anticipated for the collider detectors.
The experiments at future energy frontier hadron
colliders will have even worse problems coping with luminosities that, as
was shown in section~\ref{subsec:wishlist_lum},
should ideally rise as the square
of $\ECoM$. The problems arise from the large cross section for soft
background interactions:
\begin{equation}
\sigma^{pp}_{TOT} \simeq 200\; {\rm millibarns},
\end{equation}
rising only slightly with $\ECoM$. The average number of background
events per bunch crossing, $n_b$, is given by,
\begin{equation}
n_b = \frac{ \lum \times \sigma^{pp}_{TOT} }{f},
\end{equation}
with $f$ the bunch-crossing frequency, or, numerically,
\begin{equation}
n_b \simeq 2 \times \frac{ \lum [10^{34}\, \Lunits] }{f [{\rm GHz}]}.
       \label{pileup}
\end{equation}

  Desirable luminosities in the range $10^{35-36}\: \Lunits$ will require
major advances in detector technology and event analysis to resolve the
background event pile-up predicted from equation~\ref{pileup} and also to cope
with event-induced radiation damage to central detectors. Particular attention
will need to be paid to the radiation hardness of the central tracker and
electromagnetic calorimeter, and to fast timing, triggering and read-out of
the events.

  Equation~\ref{pileup} shows that the bunch crossing frequency, $f$, should
be made as large as is practical in order to minimize the event pile-up.
Ideally, the time between crossings,
$1/f$, should be comparable to the resolving time of the detector, which
tends to be limited to on the order of nanoseconds. (The LHC is designed
for one bunch crossing every 25 nanoseconds.) However, in practice this turns
out to be an inefficient way to produce luminosity, requiring large stored
beam currents that exacerbate the synchrotron radiation problem of
equation~\ref{ppsynch} and bring on other technical headaches.

\subsubsection{The Implausibility of Single Pass Hadron Colliders}

  An extreme example of the inefficiency of frequent bunch crossings
would be provided by any attempted design for a {\em single pass} hadron
collider that would use the linear accelerator technology developed for
$\ee$ colliders. The magnitude of the problem appears to prohibit any
serious speculation on using linear hadron colliders to extend the energy
frontier, as we now show.

  To obtain a crude scaling argument, we note that the luminosity
as a function of the bunch crossing repetition rate, $f$,
the number of particles per bunch, ${\rm N_b}$, and the transverse beam
 dimensions, $\sigma_x$ and $\sigma_y$, is given roughly by:
\begin{equation}
\lum\: \simeq\: {\rm \frac{{\mathit f}\, N_b^2}{\sigma_x \sigma_y}}
\:\propto\: {\rm  \frac{1}{{\mathit f}} \cdot
                             \frac{(P_{beam})^2}{\sigma_x \sigma_y}},
  \label{Lformula}
\end{equation}
where we haven't bothered to keep track of numerical factors depending on
the precise definitions of $\sigma_x$ and $\sigma_y$ (several different
conventions are in common use!) and have ignored effects of order unity
such as the pinch enhancement of luminosity, and the second expression
follows from the first because the average beam power, ${\rm P_{beam}}$,
is proportional to the average beam current, ${\rm P_{beam}} \propto f N_b$.
Equation~\ref{Lformula} shows that the luminosity at linear colliders falls
off inversely with the repetition rate -- at least to the extent that the
pinch enhancement can be neglected and, as a more substantial caveat,
provided that the
beam power and transverse beam dimensions are fixed. Therefore, very low
repetition rates are strongly favored and the nanosecond-scale repetition
rates desired for hadron colliders are strongly disfavored.

  To set the numerical scale, the first very speculative straw-man parameter
set for a 1 PeV linear muon collider~\cite{Zimmermann} assumes a luminosity
of $\lum = {\rm 5.4 \times 10^{35}} \Lunits$ and a repetition rate of
only $f = 3.2$ Hz. (This corresponds to a very impressive per-collision
integrated luminosity of 170 inverse nanobarns per collision!) If this
parameter set was translated to a 1 PeV hadron collider rather than a
muon collider then equation~\ref{pileup} predicts a manifestly unmanageable
$3 \times 10^{10}$ interactions per collision!

  Even if one allows for a re-optimization of parameters from this rather
extreme example, it is hard to imagine a single-pass hadron-hadron collider
with both an interesting luminosity and viable experimental conditions.
The conclusion of this subsection is then that the ultimate center of mass
energy for hadron colliders will almost certainly be attained using
circular hadron colliders and will probably be limited to a few hundred TeV.

\subsection{A 200-400 TeV Hadron Collider for the Proof-of-Plausibility
Scenario}
\label{subsec:hadron_PoP}

\subsubsection{Basic Specifications}

  The preceding subsection established that the collision energy often
assumed in VLHC studies, $\ECoM=100$ TeV, is rather close to the ultimate
energy scale possible for future hadron colliders, at -- we will guess
-- perhaps 200 to 400 TeV. It is notable that the energy jump from the LHC
($\ECoM=14$ TeV) to this energy would be about the same factor of
twenty-or-so as was planned to go from the Tevatron (1.8 TeV) to the SSC
(40 TeV). Both rate-of-progress and economy therefore suggest that it makes
eminent sense to try to reach this frontier energy in a single step. This
motivates the choice for our proof-of-plausibility scenario of a single
hadron collider after the LHC, at $\ECoM=$200-400 TeV. (The possibility
has not been excluded for an upgrade from the lower end to the higher end of
this energy range.)

  For definiteness in the overall scenario, we can also assume a 400 km
circumference for the 200-400 TeV hadron collider ring. As will
be seen later, this fits in with the size scale for a new world HEP laboratory
that also includes a muon collider.

\subsubsection{Magnets and Synchtrotron Radiation Damping}

  A 400 km circumference corresponds to an average bending magnetic field
around the collider ring of 5.3 (10.5) tesla for the assumed energy
range of $\ECoM=200$ (400) TeV.

  Because the 200-400 TeV energy
range is higher than the $\ECoM=100$ TeV value usually considered for the
VLHC, it is notable that it might give a better match between, on the one
hand, the desirable level of synchrotron radiation damping sought by those
designing high field VLHC magnets and, on the other hand, the lower magnetic
field strengths that might correspond to a cost minimum.
In fact, as a very intriguing
possibility that is more general than this scenario, it is not ruled out that
the global cost optimum for optimal synchrotron damping might be at lower
magnetic fields but higher collision energies than are currently considered
for the high field 100 TeV VLHC, i.e. the extra energy reach could conceivably
come for free! The synchrotron radiation and cost aspects of the
200-400 TeV hadron collider will now be discussed in turn.

  The levels of synchrotron radiation damping in this scenario are easily
seen to range from just slightly above, to far beyond, those encountered in
very high field magnet studies at a 100 TeV VLHC. From
equation~\ref{synchscaling}, the 5.3 T average field for the 200 TeV
scenario gives the same damping as an average field 4 times larger at
$\ECoM=100$ TeV, i.e. 21 tesla, which is slighty above the maximum
considered for VLHC studies. In contrast, an unrealistic factor of 16
in magnetic field strength at 100 TeV
would be needed to compensate for a four-fold
increase in energy, to 400 TeV, so the level of synchrotron radiation
damping in this scenario is obviously much larger. It is a subject for
further studies to determine whether the level of synchrotron radiation
at 400 TeV is desirable in, or even compatible with, any self-consistent
set of hadron collider parameters that would presumably utilize lower
beam currents, smaller spot sizes and a stronger final focus than any
current VLHC parameters.

\subsubsection{Cost Considerations}

  The basis for cautious optimism on the magnet
costs for this scenario comes from a careful study,
by superconducting magnet expert Erich Willen~\cite{Willen}, of the cost
optimum in field strength that would be obtained by assuming {\em today's}
superconducting magnet technology.

  Willen's cost evaluation is for a $\ECoM=200$ TeV hadron
collider. (As a cautionary note, he actually refers to his parameter sets
as being for 100 TeV colliders, but close inspection reveals this to be
the energy per beam rather than $E_{CoM}$.) The costing for the dipole
magnets is a careful scaling of the costs for the dipole magnets used in
the existing RHIC collider. The scaling takes into account some suggested
design modifications to increase the magnetic field and reoptimize the
magnet length, aperture and superconducting coil layout, all of which
are compatible with currently available technology. The cost vs.
magnetic field strength characteristic was found to have a rather broad
minimum reaching down to 1436 1993 \$ U.S. per tesla-meter at a
field strength of 5.7 T. In table 7 of reference~\cite{Willen},
Willen presents an estimated cost of 6 \$B for the dipole magnets in
a 200 TeV collider using two rings of 5.7 T dipole magnets with 80\%
packing, for an average bending field of 4.6 T and a circumference of 460 km.
Willen also priced the tunnelling costs for the collider ring at a little
over 0.4 \$B , costed at \$900/m after studies at the 1996 Snowmass
Workshop, i.e. a much smaller component of the total colider cost.

  To use Willen's study as a benchmark for the 200-400 TeV collider
considered here, Willen's cost minimum at a 4.6 T average field is
rather close to the required 5.3 T field for the 200 TeV collider with
a 400 km circumference, so his 6 \$ B cost estimate for today's magnet
technology is directly applicable. Improving technology in high field
superconducting magnets can then be expected to both lower the cost per
tesla-meter and raise the field strength of the cost minimum, i.e., move 
the magnetic field strength some distance
in the direction of the 10.5 T average bending
field that has been assumed for the 400 TeV energy.

  As a more quantitative statement of the progess that might be demanded
in order for 200 TeV or 400 TeV hadron colliders to become economical, it
would be helpful reduce the dipole magnet cost to about 3 \$B. This would
require a factor of about 2 or 4 reduction, respectively,
in magnet costs-per-Tesla-meter from Willen's estimate of
\$1436 /Tesla-meter. It is not unrealistic to hope that
such savings could come from economies of scale
and from a decade of technological advances in magnet components, design
and manufacture that builds on the current magnet R\&D program for the
VLHC.


\section{Muon Colliders}
\label{sec:muon}

\subsection{Circular and then Linear Muon Colliders to 1 PeV and Beyond ?}
\label{subsec:muon_limits}


\subsubsection{Switching from Electrons to Muons}

  All of the problems with TeV-scale $\ee$ colliders that were discussed
in section~\ref{sec:electron} are associated with the relative smallness
of the electron mass. The proposed technology of muon colliders aims to
solve, or at least greatly reduce, these problems by instead colliding
muons, which are leptons that are 207 times heavier than electrons.

  Replacing the mass-related problems of $\ee$ colliders, the main
problems at muon colliders arise because muons
are unstable particles, with an average lifetime of approximately
2.2 microseconds in
their rest frame. The preparation, acceleration and collision of the muon
beams must all be done quickly and the supply of muons must be replenished
often. The products of the muon decays also cause problems: the decay
electrons deposit energy all along the path of the muon beams and create
backgrounds in the detectors and, more surprisingly, the neutrinos can
cause a radiation hazard in the surroundings of the collider
ring~\cite{nurad,hemc99nurad}.

  The technology and status of R\&D on muon colliders has been covered
in detail in reference~\cite{status} and the specific issues involving
many-TeV muon colliders were examined at this workshop and form the topic
of many of the papers in these proceedings. A focal point for the studies
at this workshop was
provided by three self-consistent parameter sets for muon collider
rings, one set at 10 TeV and two sets at 100 TeV. Reference~\cite{hemc99specs}
of these proceedings discusses the parameter sets and their evaluation
at the workshop. A very significant development at the workshop was the
presentation~\cite{Zimmermann} of parameter sets for {\em linear}
muon colliders at energies ranging from 3 TeV all the way up to 1 PeV.
The general assessments on the energy reach for both circular and
linear muon colliders will now be briefly reviewed.

\subsubsection{Synchrotron Radiation Limits for Circular Muon Colliders}

  The potential energy reach for {\em circular} muon colliders appears to
hit a fairly hard limit at about $\ECoM=100$ TeV, where the radiated power
from synchrotron radiation has risen to become approximately equal to the
beam power. Additional constraints arise from beam heating due to the
quantum fluctuations in synchrotron radiation, as was pointed out in the
workshop and is discussed elsewhere in these
proceedings~\cite{Telnovsynch}.
Like the beam power, the quantum fluctuations also rise as a relatively
high power of the beam energy (both rising as energy cubed, for some
benchmark parameters -- c.f. equation~\ref{ppsynch}
for hadron colliders) -- and this further
pins the ultimate potential for circular muon colliders down to the 100 TeV
energy scale. See reference~\cite{hemc99specs} for further discussion on
the limits imposed by synchrotron radiation.

\subsubsection{Single Pass Muon Colliders}

  Following the historical path of $\ee$ colliders, muon collider energies
above 100 TeV can be contemplated by switching to the technology of
{\em linear} colliders, as was shown by Zimmermann~\cite{Zimmermann}. All
of Zimmermann's parameter sets -- at 3, 10, 100 and 1000 TeV -- require
specified ``exotic'' technologies both for preparing the muon beams and
for acceleration, although these are not implausible at first reading
by this author, who is admittedly not particularly knowledgable
about linear colliders; see reference~\cite{hemc99specs}
for further discussion.
(Zimmermann's provocative parameter sets clearly need further review
by people more expert than this author.) 

  It is a remarkable feature of the progression in parameters that the final
parameter set, at $\ECoM=1$ PeV, appears to be not so much more ``exotic''
or less technologically plausible than the initial parameters at 3 TeV.
This justifies the inclusion of a PeV-scale linear muon collider as the
last collider in the proof-of-plausibility scenario, where it can benefit
from about three decades of R\&D to develop and refine the necessary
technologies. Conversely, circular muon colliders have been favored over
linear colliders up to the 100 TeV scale, where their technologies seem
better established.


  Continuation of linear muon colliders to even beyond the PeV scale
has not yet been rigorously excluded, although the
technological challenges would obviously be formidable. For example,
the final focus design is far removed from anything we can seriously
contemplate today. Also, the 2 linacs for even a 1 PeV
collider are each already 500 km long for an assumed accelerating
gradient of 1 GV/m, giving a maximum depth of 5 km below a spherical
Earth. (See reference~\cite{hemc99nurad} for an illustration of the
geometry.)

  A 10 PeV collider would need either 10 times the tunnel length or 10
times the accelerating gradient of the 1 PeV example, or some compromise
between these parameters. However,
the maximum depth below the Earth's surface goes as the square of
tunnel length, so trying to lengthen the tunnel quickly gets one
into hot lava!
The barriers against increasing the gradient to 10 GV/m have more
loopholes. Gradients of 1 GV/m are already pushing the material
limits for any known surfaces in the accelerating structures, even at
the highest frequencies people consider, but the solution to this might
plausibly come from exotic acceleration schemes using plasmas driven by
lasers. Various experiments to test such acceleration schemes are already
underway.

\subsection{Muon Colliders for the Proof-of-Plausibility Scenario}
\label{subsec:muon_PoP}

  This subsection provides discussion that is specific to the 3 muon
colliders of table~\ref{timeline}, at $\ECoM=4$ TeV, 30-100 TeV and
1 PeV. In order to spell out the entire progression, however, we begin
by discussing a neutrino factory -- a simpler accelerator that is not
actually a collider but which would likely serve as an important staging
point towards the construction of the 4 TeV muon collider that is the
first $\mm$ collider entry in table~\ref{timeline}.


\subsubsection{Leading-in with a Neutrino Factory Muon Storage Ring}

  Because no muon collider has yet been built, a convincing demonstration
of muon collider technology would be prudent before investing in an energy
frontier collider. A neutrino factory would afford this demonstration
while providing much useful and complementary physics to the colliders.

  A neutrino factory is a simpler, non-colliding muon storage ring that
would be optimized for the collimated decay of muons into one or more intense
neutrino beams aimed at neutrino physics experiments. As an aside on the
choice of beam energy for the neutrino factory, current studies are
concentrating mainly on neutrino factories with beam energies of 20 to
50 GeV that are optimized for a particular range of neutrino oscillation
studies. However, neutrino factories at higher beam energies of perhaps
100--200 GeV are more optimal for the complementary high-rate experiments
that study neutrino interactions and also for some other neutrino
oscillation scenarios. This option for neutrino factories at higher energies
might also be further considered since it would better test the
acceleration technology needed for high energy muon colliders and
should also be easier to upgrade to such a collider.

  In the straw-man scenario presented here, the design of the neutrino factory
demonstration machine would be made compatible with an upgrade to a
site-filling 4 TeV muon collider, to occur immediately on completion
if not beforehand. The neutrino factory provides an ideal intermediate step in
the construction of a 4 TeV collider because it would be a valuable
HEP facility in its own right if, for some reason, the continued
upgrade to a collider was found not to be technologically feasible.
It can therefore be built before a final decision has been made on
the collider technology, and its construction can better inform that
decision.

  As another nice feature of this staging scenario, the experimental
investments and the incremental physics from neutrinos is not lost with
the upgrade to a collider.
Quite the contrary, it turns out~\cite{nuphysref} that much of the
high-rate neutrino physics can anyway be performed even better in parasitic
running at a multi-TeV collider than in
a dedicated lower-energy neutrino factory.
The schedule for the proof-of-plausibility scenario would require a
decision by 2007-8 on whether or not to proceed to a site-filling muon
collider. By this time, the decision could presumably be
guided by preliminary results from 14 TeV CoM proton-proton
collisions at the LHC.

\subsubsection{A 4 TeV Muon Collider}

  Four TeV was the muon collider energy chosen for the design
study~\cite{Snowmass} presented at the Snowmass'96 workshop.
However, it has since become more widely appreciated that
neutrino radiation from the collider ring is a concern at these
energies~\cite{nurad}, so the muon beam current would likely be
limited to more than an order of magnitude below the Snowmass'96
parameters. The reduced current would anyway save on the cost and
power of the proton driver, which
might have been considerable for the Snowmass'96 specifications.

  Reference~\cite{epac98} contains a self-consistent parameter set
for a low-current 4 TeV muon collider that is appropriate for the
proof-of-plausibility scenario given here. The parameter set has a
luminosity of $\lum=6 \times 10^{33}\: \Lunits$
and an average neutrino radiation dose in the
plane of the collider that is below one-thousandth of the 1 mSv/year
U.S. federal off-site limit.

\subsubsection{The Highest Energy Circular Muon Collider, at 30-100 TeV}


  The next step up in energy is assumed to carry all the way to the
highest feasible energy for a circular machine, which we assume
to be in the range $\ECoM=30$-100 TeV.

  In order to consolidate resources and minimize the overall cost of
the scenario, it is sensible that the 30-100 TeV $\mm$ collider should
be constructed in the same laboratory whose 400 km boundary is defined
by the 100-200 TeV hadron collider of the preceding section.
A boundary of such a size will
anyway be dictated by the requirements of neutrino radiation, as
is discussed in detail elsewhere in these proceedings~\cite{hemc99nurad}.
Briefly, the radiation disk that is emitted in the plane of the
collider ring would rise to approximately 300 meters above the Earth's
surface for such a boundary and in the approximation of a spherical
Earth. This is assured to be far enough above any structures that no
practical off-site radiation hazard would remain, as is discussed
further in~\cite{hemc99nurad}.


  The high end of the energy range, $\ECoM=100$ TeV, corresponds to two
of the straw-man parameter sets for this workshop -- see the parameter
table in reference~\cite{hemc99specs} -- while parameters for the low
end, 30 TeV, can be estimated by interpolating between the 10 TeV and
100 TeV straw-man parameter sets in reference~\cite{hemc99specs}. As we
learned in the workshop~\cite{Telnovsynch}, beam heating effects from
synchrotron radiation push the parameters in the direction of a
lower average bending field than the 10.5 T assumption of the
workshop's parameter sets. A sensible value might, e.g., be half of this,
i.e. a 5.3 Tesla average for a 200 km circumference at 100 TeV,
as was indicated in reference~\cite{hemc99specs}. Such
effects should be much less important if the maximum collider energy
turns out to be limited towards the lower end of the energy range under
consideration. A 30 TeV collider would presumably still use as high an
average bending field as is practicable in order to maximize its
luminosity, e.g., perhaps as high as a 10.5 T average. This would
correspond to the much smaller circumference of only 30 km.

  As for the hadron collider, bending magnets are likely to be the
major cost component of this $\mm$ collider. The collider ring magnets
should be much cheaper for a 30-100 TeV muon collider than for the
200-400 TeV hadron collider as they would require an order of magnitude
less tesla-meters of total bending: the beam energy is several times
lower and there is a further factor-of-two saving because only one ring of
magnets is needed instead of two (counter-rotating beams of opposite
charges can share the same magnet ring). Indeed, a presentation at this
workshop by Mike Harrison~\cite{Harrison} suggested a total magnet cost
of only about 400 million dollars
for the collider ring magnets in the 10 TeV parameter
set. This is very encouraging to the extent that it can be scaled up
to higher energy muon colliders (although see Harrison's caveats
regarding such a scaling). Instead, the magnets for the
muon acceleration are likely to be a much larger cost component than
the collider ring magnets, with perhaps several times the total
tesla-meters of bending as well as additional costs associated with the
need to transport large momentum spreads. (See reference~\cite{hemc99specs}
of these proceedings for a more detailed discussion.) The design of the
magnets for both the acceleration and collider rings will also need to
cope with the energy deposited from decay electrons and sychrotron
radiation, and this should also feed down into more expensive magnets
than at hadron colliders.
The upshot of the discussion in this paragraph is that the cost of
the 30-100 TeV muon collider might plausibly be similar to that of the
200-400 TeV hadron collider in the scenario. There would probably
also be a substantial overlap in the magnet technologies that drive
the costs of the two colliders.

  It is noted that, as a follow-up to the workshop's
parameter sets~\cite{hemc99specs}, new
$\mm$ collider parameter sets are currently being generated~\cite{epac2000}
at the $\ECoM=30$ TeV and 100 TeV lower and upper limits of the range
specified here.

\subsubsection{The 1 PeV Linear Muon Collider}

  The final collider in table~\ref{timeline} is a 1 PeV linear $\mm$ collider.
This item in the proof-of-plausibility
scenario simply defers to an expert in linear colliders by assuming
the 1 PeV parameters that were presented elsewhere
in this workshop by Zimmermann~\cite{Zimmermann} and have already been
discussed previously in this paper.
(If a change to Zimmermann's scenario were to be guessed at, it might be
in the direction of more frequent bunches with smaller emittances but
fewer muons per bunch. Such a refinement appears to move towards the
anticipated potential capabilities of emittance reduction using the
proposed method of optical stochastic cooling~\cite{Zholents}.)

  Zimmermann does not specify a linac length or accelerating gradient, so
this scenario makes the additional assumption of an accelerating
gradient of 1 GV/m. This corresponds to
a 500 km total length in each of the 2 linac tunnels.
This gradient is rather ambitious, as was pointed out in the preceding
subsection. The new laboratory site should have provision for a 1000 km
long linear tunnel centered in the laboratory site, as is shown in
figure~\ref{worldlab}.

  In order for the beam acceleration to have some hope of an
acceptable cost scaling with energy, Zimmermann assumes an electron-drive
beam technology for the acceleration, such as is now under development for
the CLIC linear $\ee$ collider. To be affordable at 1 PeV, this technology
will need to reduce the cost-per-unit-length of the linac by more than
an order of magnitude over the more conventional klystron-driven technology,
to below \$ 10 000/meter, corresponding to a total linac cost below \$ 10B.

  The luminosity of Zimmermann's 1 PeV parameter set is
$5.4 \times 10^{35}\: \Lunits$. This corresponds to only about half a
unit of R in integrated luminosity per $10^7$ second accelerator year
(see section~\ref{subsec:wishlist_lum}) and so falls squarely on top of
the ``borderline'' luminosity assignment of equation~\ref{Lborderline}:
\begin{equation}
\lum^{\rm borderline} {\rm [\ECoM = 1\: PeV]} =
             {\rm  5 \times 10^{35} \Lunits}.
\end{equation}

  The 1 PeV linear $\mm$ collider is the farthest extrapolation in time,
energy and technology
of all the colliders in the proof-of-plausibility scenario, and its overall
feasibility and choice of parameters also have the largest uncertainties.
Even so, reference~\cite{Zimmermann} was really only a first look at linear
$\mm$ colliders and there is presumably much that can be done to further
assess and develop this possibility with studies that wouldn't be too
time-consuming. The interested reader is encouraged to take a further
look!

\section{Assessment of Challenges and Prospects for the
Proof-of-Plausibility Scenario}
\label{sec:scenario}

  The individual challenges for each of the $\ee$, pp, and $\mm$ collider
technologies in the proof-of-plausibility scenario have been briefly
addressed in the preceding sections. The three subsections in this
section now discuss the global features: the common technologies,
overall costs and the potential physics rewards.

\subsection{Technological and Logistical Requirements for the
                                            Overall Scenario}
\label{subsec:scenario_tech}

\subsubsection{Common Technologies}

\begin{table}[htb!]
\caption{
Technology overlaps for future colliders at the energy frontier.
}
\begin{center}
\begin{tabular}{|c|ccc|}
\hline
Technology & pp & ${\rm e^+e^-}$ & $\mu^+\mu^-$ \\
\hline
\multicolumn{1}{|l|}{{\bf magnets:}}  & & & \\
SC conductors         & Y & n & Y \\
cost reduction        & Y & n & Y \\
v. high B dipoles     & Y & n & Y \\
v. high B quads.      & Y & n & Y \\
heat removal          & Y & n & Y \\
FFAGs                 & n & n & Y \\
\multicolumn{1}{|l|}{{\bf acceleration:}}  & & & \\
high grad. SC rf      & n & Y & Y \\
high grad. normal rf  & n & Y & Y \\
drive-beam            & n & Y & Y (linacs) \\
laser-driven plasma   & n & ? & far future? \\
\multicolumn{1}{|l|}{{\bf other hardware:}}  & & & \\
OSC                   & ? & n & Y (linacs) \\
beam diagnostics \& feedback     & Y & Y & Y  \\
active magnet movers            & n & Y & Y (linacs) \\
\multicolumn{1}{|l|}{{\bf design and simulations:}}  & & & \\
hard final focus design  & ? & Y & Y \\
beam stability studies   & Y & Y & Y \\\hline
\end{tabular}
\label{tech}
\end{center}
\end{table}

  It has been noted in the preceding sections that pp and circular $\mm$
colliders share the technology of high field bending magnets as their cost
drivers. In fact, it is true more generally that the technologies for progress
in the three accelerator types are very much intertwined, as is itemized
in table~\ref{tech}.

  The considerable overlap serves to remind us of the common future of
the field and to provide further impetus for cooperative technical studies
between experts in each of the 3 accelerator types, and the technological
health and vigor of each one of these accelerator
types will trickle down to affect the others.

  All of these technologies will require considerable R\&D if a viable
rate of progress towards the energy frontier is to be maintained.
This implies a concerted and, probably, expanded commitment of resources
to accelerator technologies, as was already pointed out several years ago
by the Nobel laureate experimental physicist Samuel C. C. Ting~\cite{Sci Am}:
``We need revolutionary ideas in accelerator design more than we need
theory. Most universities do not have an accelerator course. Without
such a course, and an infusion of new ideas, the field will die.''

\subsubsection{The Desirability of a New World HEP Laboratory}

\begin{figure}[t!] %
\centering
\includegraphics[height=3.0in,width=4.0in]{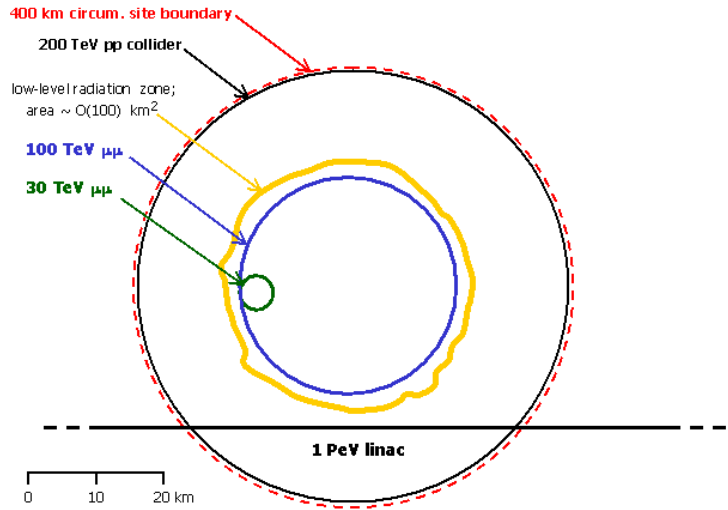}
\caption{
The ultimate high energy physics laboratory ?
}
\label{worldlab}
\end{figure}

\begin{figure}[t!] %
\centering
\includegraphics[height=3in,width=3in]{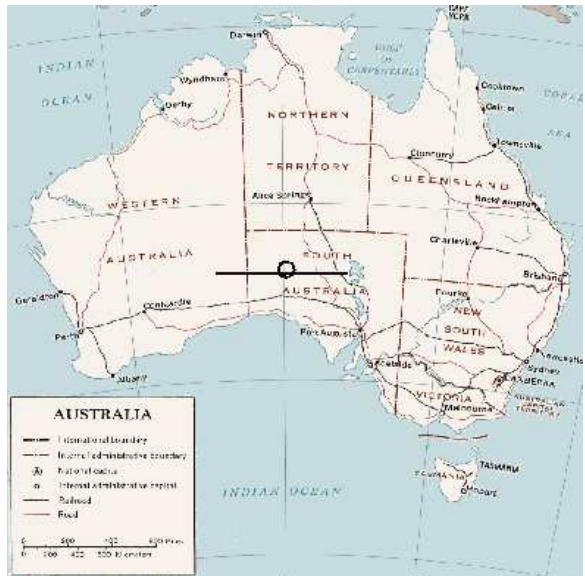}
\caption{
  An example to illustrate the size of a HEP laboratory
with a 400 km site boundary circumference. A circle of this
diameter has been drawn in the Great Victoria Desert
(just above the ``A'' in the label ``South Australia''),
showing that the outline of a laboratory of this size
would even be visible on a map of Australia.
The 1000 km long strip of land for the 1 PeV linear collider
would perhaps not be included inside the original laboratory
boundary. The choice
of country and positioning of the site are for illustration
only.
}
\label{labAustralia}
\end{figure}

  As well as the technological overlap, the proof-of-plausibility scenario
assumed that the 200-400 TeV hadron collider and the final circular and
linear $\mm$ colliders would be housed in the same new world HEP laboratory.
An ``artist's conception'' example illustration of the layout for this
scenario is shown in figure~\ref{worldlab} and its large size scale has
been emphasized in figure~\ref{labAustralia} by placing it somewhat
arbitrarily on a map of Australia.

  The choice of an appropriate large laboratory site would obviously be a
major project involving a considerable amount of research followed by detailed
political negotiations. Site selection would involve the optimization of many
factors and the satisfaction of several requirements besides
isolation~\cite{Colin}. For example, the site should be in a politically and
economically stable country and should have ready access to an industrial
base and resources such as power, transport and cooling water.
Besides Australia, other candidate regions for a site would obviously
include the U.S.A., Canada and several parts of Northern Europe or Asia.
The example of figure~\ref{labAustralia}, showing the site in an unpopulated
desert region of Australia, would presumably use closed cooling loops to
conserve water. It recalls the discussion of around 1980 on
the ``desertron'', which was the original popular name for a mooted
energy frontier hadron collider that then evolved into the SSC project.

  The consolidation of resources into a single new HEP laboratory should
be more generally beneficial in all scenarios that envisage more than
one collider extending the high energy frontier, not to mention that even
a single such collider will anyway be too big to fit on existing laboratory
sites. This would also
help financially by avoiding duplication of laboratory operating costs --
an aspect that will be covered in more detail in the following subsection.

\subsection{Could the Scenario be Affordable ?}
\label{subsec:scenario_cost}


  For a plausible funding scenario, note that the straw-man scenario includes
5 energy-frontier colliders beyond the LHC (1 $\ee$, 1 pp and 3 muon
colliders) over a time period of about 40 years. The following suggested cost
goals appear to be plausible and correspond to an assumed level of worldwide
construction funding on these machines of 1 B\$/year over these 40 years:
\begin{enumerate}
   \item  10 \$B for the combined cost of the $\ee$ collider, the neutrino
                factory and the 4 TeV muon collider
   \item  8 \$B for the 200-400 TeV hadron collider
   \item  10 \$B for the final circular muon collider, at 30-100 TeV
   \item  12 \$B for the 1 PeV linear muon collider.
\end{enumerate}

  The 10 \$B cost combined cost for the first machines in the scenario can
be estimated with slightly more confidence than the others and appears
difficult but plausible. The cost drivers contributing to the rest of these
guessed figures have been at least touched on in the preceding sections but
the potential reasonableness or otherwise of such cost goals is clearly a
subject requiring much more careful and detailed study. Ideally, a global
cost assessment of funding scenarios would benefit from explicit funding
algorithms that have been benchmarked to recent large accelerator projects
such as the SSC, LEP, HERA, RHIC and the LHC, and then peer-reviewed and
continually refined by the HEP community as our technical understanding
improves.

  Besides the costs of the accelerators themselves, additional costs would
include accelerator R\&D and the commissioning and operating costs of the new
world HEP laboratory. The laboratory would presumably cost several billion
dollars to set up, although this might reasonably be partially or fully covered
by a one-off contribution from the host country. Operating costs and upkeep
for such a large laboratory might amount to several hundred million dollars
per year, considering that the world's largest current laboratory, CERN, has
an annual budget in the range of half a billion dollars.
Electricity for the colliders would be a significant part of the laboratory's
operating cost; to set the scale, operating for a $10^7$ second accelerator
year with a total wall-plug power of 1 Gigawatt would cost 140 million dollars
at an assumed rate of 5 cents per kilowatt hour.

  All-in-all, the potential viability of the proof-of-plausibility scenario
assumes something around 2 \$B per year as the world-wide HEP budget devoted
to the high energy frontier. This can be compared to the total funding for
HEP in the U.S. alone~\cite{Drell}, which has bounced around at an average
slightly below 1 \$B per year since the early 1960's but has risen as high
as 1.2 \$B in 1970 and 1.4 \$B in 1992, at the height of SSC funding. (These
figures are in FY 1995 U.S. dollars.) This indicates that the scenario could
not easily be funded by one country acting alone but would instead require the
combined commitments of all of the U.S., continental Europe, Britain, Japan
and other, smaller contributors.

  To summarize the cost discussion, the obvious answer to the question in
the section title is ``not easily''.
The cost constraints will be extremely challenging in the straw-man collider
scenario presented here, or in any other that holds to the historical
precedent for progress along the Livingston plot. The costs of each machine
will need to be agressively minimized and, even so, the field will need to
make a united and convincing case for concerted world-wide funding at a
level that is at least comparable to the historical norm for each of the
contributing countries.

\subsection{Benchmarking the Experimental Potential of Future Colliders} 
\label{subsec:scenario_phys}

  The preceding discussion in this section has addressed the ``pain''
involved in keeping to the Livingston curves for accelerator progress.
We now turn to the ``gain'' that would make all the effort worthwhile.
If the future colliders covered in this proof-of-plausibility scenario
turned out to be indeed feasible and were constructed, it would extend our
energy reach for elementary particles from the current 300 GeV flagship
in energy reach (the 1.8 TeV proton-antiproton Tevatron) all the way up
to the 1 PeV energy scale.

  The true significance that such an experimental advance would have
is unknowable until it happens
since it depends on what we find and how successful we are in
tying it in to our theoretical understanding of the Universe. All we
can do for now is extrapolate from past experience in particle
physics. Such an advance of $3 \frac{1}{2}$ decades in energy reach
can be judged against these benchmarks:
\begin{itemize}
  \item   the past and present
colliders on the Livingston plot span from about 1 GeV to 300 GeV.
These $2 \frac{1}{2}$ energy decades of collider experiments
have been sufficient to revolutionize our knowledge of the
elementary constituents of our universe. The final span of
6 energy decades would be an increase on this by a factor
of 2.4, and it would be pessimistic to not predict that this
would again revolutionize our level of understanding.
  \item   those elementary particles in figure~\ref{particlechart}
that are not massless have a mass spectrum extending from the electron,
with $m_e \simeq 5.11 \times 10^{-4}$ GeV, to the top quark,
with $m_t \simeq 1.75 \times 10^{2}$ GeV. (We exclude for the
moment the very preliminary and indirect experimental evidence for
non-zero neutrino masses that, if confirmed, would be much lower.) Therefore,
the additional $3 \frac{1}{2}$ energy decades could potentially
broaden the known spectrum of elementary particle masses from
the current $5 \frac{1}{2}$ decades in mass to 9 decades, a substantial
increase of more than 60\%.
  \item   $3 \frac{1}{2}$ decades in energy reach would explore more than
20\% of the log-energy gap from our
current experimental reach all the way up to the Planck mass scale,
at $10^{19}$ GeV. The Planck scale is defined by where quantum
gravitational effects necessarily become important enough that even
the framework for our current theories no longer makes sense and
some theory that would surely be a close approximation to the
Theory of Everything would be required to describe the physical
processes. By biting off a significant fraction of this log-energy
span we would presumably be giving ourselves a good shot at such an
elevated level of understanding of our Universe.
\end{itemize}

\section{Conclusions}
\label{sec:concl}

  This paper has reviewed the past and future importance of accelerators for
understanding our cosmos and its elementary constituents. The prospects for
future $\ee$, pp and $\mm$ colliders were also reviewed and a
proof-of-plausibility scenario was presented, incorporating the 5 plausible
future colliders in table~\ref{timeline},
that is able to hold to the historical rate of
progress in the log-energy reach of hadron and lepton colliders and to reach
the 1 PeV constituent mass scale by the early 2040's.

  While the challenges to a further half century of concerted progress on
energy frontier colliders are great, the potential rewards are grander.
Experimental discoveries at colliders would be expected to feed further
advances in theoretical areas such as, e.g., string theory and, to reciprocate,
any such theoretical advances would then motivate and inspire continued
advances in colliders and collider experiments. The side-by-side progress
of experiment and theory have the common and lofty goal of uncovering the
long sought after ``Theory of Everything'' -- the sum total of the elementary
entities and organizing principles that underly the structure and processes
of our physical Universe. Such an immortal pillar of knowledge and
understanding, if attained, could justifiably be regarded as the greatest
scientific achievement in all of human history, no less!

  The high stakes and long-term nature of this scientific endeavor underly
the wisdom of devoting some small fraction of our energies towards better
understanding the possible options and required technologies for the years
and decades to come. Proof-of-plausibility scenarios such as the one presented
in this paper can guide this planning. However, they encompass diverse and
often speculative areas of expertise and so will necessarily start out poorly
informed and in need of refining in the furnace of scientific peer review,
to then be augmented by alternative scenarios and either developed further
or else disgarded as unrealistic.

  Such a spirit of friendly and cooperative problem-solving, constructive
criticism and model-building was a hallmark of this workshop. HEMC'99 was
restricted to exploring the technologies and collider physics of many-TeV
muon colliders but planning is underway for a follow-up study and workshop,
in the Summer and Fall of 2001,
that will explore the long-term prospects for all types of energy frontier
colliders and the physics processes they might illuminate.



\section{Acknowledgements}

\end{document}